\documentstyle[prl,aps]{revtex}
\input epsf
\tighten

\begin{document}

\draft
\title{Field configurations with half-integer angular momentum 
in purely bosonic theories without topological charge}


\author{
Tanmay Vachaspati}
\address{
Physics Department, 
Case Western Reserve University, 
Cleveland OH 44106-7079.
}


\wideabs{

\maketitle


\begin{abstract}
\widetext

It is shown that purely bosonic field theories can have 
configurations with half-integral angular momentum
even when the topological magnetic charge of
the configuration vanishes. This result is applicable whenever
there is a non-Abelian gauge theory with particles that transform
in the fundamental representation of the non-Abelian symmetry
group.

\end{abstract}


}

\narrowtext

The occurrence of half-integer angular momentum configurations 
in purely bosonic theories has been known and studied for over 
two decades in the context of magnetic monopoles \cite{rjcr,phth}. 
The basic idea is described quite simply as follows. Let us suppose 
that a theory has magnetic monopoles having asymptotic magnetic field
$$
{\bf B} = {{\bf M} \over {g}} {{{\hat {\bf e}}_r} 
                                 \over {r^2}} \ ,
$$
where ${\bf M}$ is the matrix-valued magnetic charge and $g$
is a coupling constant. Consider the monopole in the presence of
some electric charges in the theory. Then the angular momentum in 
the gauge fields is
$$
|{\bf J} |= \biggr | \int d^3x {\rm Tr} [ {\bf r} \times
              ({\bf E} \times {\bf B} ) ] \biggr | 
= ~ \bigr | - \sum_i m_i e_i  \bigr | \ ,
$$
where, $m_i$, $e_i$ are the magnetic and electric charges and
the different possible types of charges are labelled by $i$. 
The Dirac quantization condition permits
the product of magnetic and electric charge to be half-integral
and hence the angular momentum can be half-integral even with purely 
bosonic particles present in the original theory.

The best known example where half-integral spin is obtained is in
the case of the 't Hooft-Polyakov monopole \cite{th,ap} with an
additional scalar field transforming in the fundamental representation
of $SU(2)$ \cite{rjcr,phth}. There the monopole
only has a single type of charge ($i=1$) and 
$$
J = - m e  = {1\over 2} \ .
$$
This phenomenon is described by saying that the $SU(2)$ isospin degree
of freedom has led to the (half-) spin of the monopole. The spin-statistics
relation is valid for the monopole \cite{ag} and so the monopoles are
fermionic objects in a theory of bosons.

Spin from isospin has been considered for the more complicated $SU(5)$
monopoles by Lykken and Strominger \cite{ls} and we shall now discuss
and generalize
their analysis. The idea is to calculate $\sum m_i e_i$ for various
monopoles and electric charges. For the monopoles formed in
\begin{equation}
SU(5) \rightarrow [ SU(3)\times SU(2)\times U(1) ]/Z_6 = K
\label{model}
\end{equation}
there are four magnetic charges ($i=1,...,4$) corresponding to the
four diagonal generators of $K$. These are $\lambda_3$ and $\lambda_8$ 
of $SU(3)$, $\tau_3$ of $SU(2)$, and, $Y$ of $U(1)$. The magnetic charges
can be chosen to be:
$$
m_1 = 0\ ,\ m_2 = {{n_8} \over {\sqrt{3} g}}\ , \ 
m_3 = {{n_3} \over {2 g}}\ , \ 
m_4 = {{-1} \over {2g}} \sqrt{5 \over 3} n_1 \ ,
$$
where, if $n$ is the winding of the monopole,
$$
n_8 = n + 3k \ , \ \ n_3 = n +2l\ ,
\ \  n_1 = n  \ ,
$$
where $k$ and $l$ are arbitrary integers. The reason for the freedom
in choosing $k$ and $l$ is that $m_2$ and $m_3$ are actually
$Z_3$ and $Z_2$ charges respectively and the addition of $3k$ and $2l$
to $n_8$ and $n_3$
does not make any difference to the topological winding of the monopole.
However, there are minimal values of $n_8$ and $n_3$ which are
$$
n_{8,{\rm min}} = n ({\rm mod} \ 3) \ , \ \ 
n_{3,{\rm min}} = n ({\rm mod} \ 2) \ .
$$

A scalar field $H$ transforming in the fundamental representation of
$SU(5)$ has five components (labelled by $h=1,2,...,5$) which have the 
following electric charges:
\begin{eqnarray*}
e_1^h &=& {{g} \over {2}}\pmatrix{1\cr -1\cr 0\cr 0\cr 0\cr} \ ,\  
e_2^h = {{g} \over {2\sqrt{3}}}\pmatrix{1\cr 1\cr -2\cr 0\cr 0\cr}\ , \\ 
e_3^h &=& {{g} \over {2}}\pmatrix{0\cr 0\cr 0\cr 1\cr -1\cr}\ , \ 
e_4^h = {{g} \over {\sqrt{15}}}\pmatrix{1\cr 1\cr 1\cr -3/2\cr -3/2\cr}\ . \\
\end{eqnarray*}
Then, we find
$$
J^h = - \sum_i m_i e_i^h =
\pmatrix{(-n_8+n_1)/6\cr (-n_8+n_1)/6\cr (+2n_8+n_1)/6\cr 
                        (-n_3-n_1)/4\cr (+n_3-n_1)/4\cr} \ .
$$
By inserting appropriate values of $n_8$, $n_3$ and $n_1$ for each
winding $n$, one can determine whether the monopole can have half-integral
spin. For $\pm n = 1,2,3,4$, and, $6$, the monopoles are stable \cite{gh}, 
and for minimal values of $n_8$ and $n_3$, half-integral spin is always 
possible.

This shows that one can get spin from isospin for $SU(5)$ monopoles.
But we are interested in spin from isospin in the absence of magnetic
charge. For this, consider the unit winding monopole already analysed
in \cite{ls}. This has $n_1=1$, $n_3=1$, $n_8=1$ and so
$$
J_{+1}^h = \pmatrix{0\cr 0\cr +1/2\cr -1/2\cr 0\cr}\ .
$$
Also consider the winding $-1$ monopole with $n_8=-1$, $n_3=+1$, $n_1=-1$.
Then,
$$
J_{-1}^h = \pmatrix{0\cr 0\cr -1/2\cr 0\cr +1/2\cr}\ .
$$
Next consider the situation where a winding $+1$ and winding $-1$ are both
present. Then,
$$
J_{+1-1}^h = - \sum_i (m_i+m_i') e_i = J_{+1}^h + J_{-1}^h =
\pmatrix{0\cr 0\cr 0\cr -1/2\cr +1/2\cr}\ .
$$
So the spin can indeed be half - even though the net topological 
magnetic charge of the system vanishes. The underlying reason for this
is that the $SU(3)$ and $SU(2)$ gauge fields contribute to the angular
momentum but the topological charge only depends on the $U(1)$ charge.

\begin{figure}[tbp]
\caption{\label{vofr}
A schematic plot of the
interaction potential of the monopole with $(n_1,n_3,n_8)=(1,1,1)$
and the antimonopole with $(n_1,n_3,n_8)=(-1,+1,-1)$. 
}

\

\epsfxsize = \hsize \epsfbox{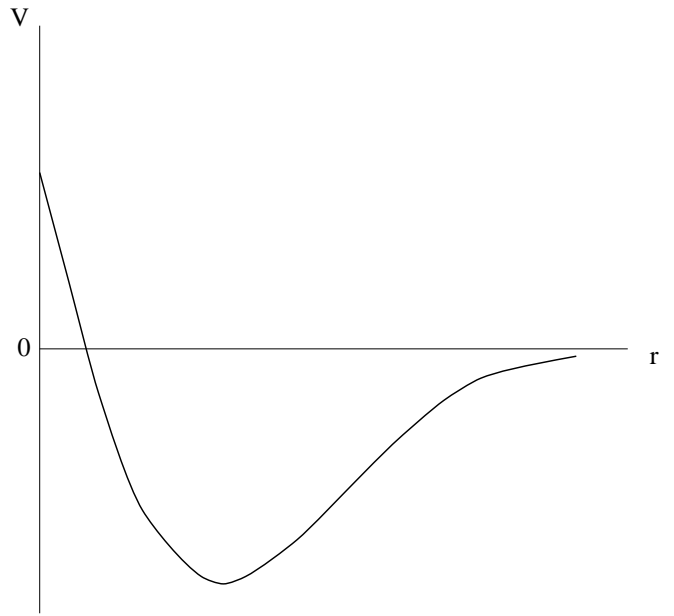}
\end{figure}

The monopole and antimonopole pair with spin 1/2 also form a bound 
state.  This can be seen by considering the interaction potential 
between an $(n_1,n_3,n_8)$ and an $(n_1',n_3',n_8')$ monopole
\cite{gh,hltv}:
\begin{eqnarray}
V(r) = {1 \over {4\alpha r}} [
       & n_1 n_1'& {\rm Tr}(Y^2) (1-e^{-\mu_0 r}) + \nonumber \\
       & n_3 n_3'& {\rm Tr}(\tau_3 \tau_3' ) (1-e^{-\mu_3 r}) + \nonumber \\
       & n_8 n_8'& {\rm Tr}(\lambda_8 \lambda_8 ') (1-e^{-\mu_8 r}) ] 
\label{intenergy}
\end{eqnarray}
where $r$ is the monopole-antimonopole separation, $\alpha = g^2/4\pi$,
the primes on $\tau_3 '$ and $\lambda_8 '$ denote that these could be 
in orientations that are different from $\tau_3$ and $\lambda_8$, and
the parameters $\mu_0$, $\mu_3$ and $\mu_8 = \mu_3 /2$ are masses of
adjoint scalar fields in the model in (\ref{model}). We are interested
in the case $\mu_0 << \mu_8$. Furthermore, the monopoles we are considering
have $\lambda_8 ' = \lambda_8$, $\tau_3' =\tau_3$, $n_1'=-n_1 =1$, 
$n_3'=+n_3 =1$ and $n_8'=-n_8 =1$. The normalization used in (\ref{intenergy})
is: ${\rm Tr}(Y^2) = 5/6$, ${\rm Tr}(\tau_3^2) = 1/2$, and 
${\rm Tr}(\lambda_8^2) = 2/3$. This leads to:
\begin{eqnarray*}
V(r) =  {1 \over {4\alpha r}} [
       -&{5\over 6}& (1-e^{-\mu_0 r}) + \\
        &{1\over 2}& (1-e^{-2 \mu_8 r}) -
        {2\over 3} (1-e^{-\mu_8 r}) ] \ . 
\end{eqnarray*}
For $r$ near zero, a Taylor expansion gives:
$$
V \rightarrow {1\over 3} \biggr (\mu_8 - {5\over 2}\mu_0 \biggr )
   - {1\over 3} \biggr (\mu_8^2 - {5\over 4} \mu_0^2 \biggr ) r + ...
$$
Therefore $V$ is positive and decreasing near the origin. As 
$r \rightarrow \infty$, $V$ approaches zero from below since the
exponentials can be neglected at large $r$ leading to
$V \sim -1/r$. This explains the schematic shape of the 
potential shown in Fig. \ref{vofr}. The presence of a minimum in the
potential shows that the monopole and antimonopole can form a bound state. 
(This is similar to the monopole-antimonopole bound
state found by Taubes \cite{taubes} in an $SO(3)$ model.) In
the presence of a suitable electric charge, this bound state 
carries half-integral angular momentum.

It is helpful to describe the half-integral values of the angular
momentum somewhat differently. If we
had taken $n_3 =-1$ for the winding $-1$ monopole, we would have obtained
the expected result that the angular momentum $J_{+1-1}^h$ is zero.
This may be made more explicit by writing $n_3=+1$ for the winding $-1$
monopole as  $-1 +2$, and then the half-integral
contribution to $J$ comes from the interaction of the $+2$ part of
$n_3$ with the electric charge. Indeed it can be seen that by changing
$k$ (and similarly $l$) by $1$, we can change the spin of certain
components by units of $1/2$. In other words, purely gluonic excitations
of the monopole, having nothing to do with topological charge,
can change the spin by half an integer. 

It is also possible to write down localized field configurations (not 
solutions) in an $SU(2)$ model which carry half-integral angular momentum. 
Consider the $SU(2)$ gauge model with a scalar field $\Phi$ that transforms 
in the fundamental representation of $SU(2)$. (Note that there is no adjoint 
scalar field present.) Then the gauge field configuration of the 
Bogomolnyi-Prasad-Sommerfield (BPS) monopole \cite{b,ps}:
\begin{equation}
A_i^a = {{\epsilon_{aij}} \over {gr}} \biggl [
        1-  {{C r} \over {{\rm sinh}(Cr)}} \biggr ] {\hat {\bf x}}_j \ ,
\label{bpsa}
\end{equation}
where $C$ is any constant, together with a quanta of the $\Phi$ field, has 
angular momentum 1/2 for every value of $C$. This follows because we know
that the gauge fields of the BPS monopole and a quanta of $\Phi$ carry
half-integer angular momentum \cite{rjcr,phth}. However, the unbroken
$SU(2)$ model has no monopoles and the gauge field (\ref{bpsa}) has
no topological charge.
Another way of stating this result is that non-Abelian theories
may have embedded monopole configurations \cite{mbmbtv} which do not
carry any topological charge but can still yield half-integral angular
momentum in the presence of a suitable electric charge. 

A potential application of these results may
be to QCD where we have $SU(3)_c$ gauge fields and hence can consider
configurations of the form given in eq. (\ref{bpsa}) (within an $SU(2)$
subgroup of $SU(3)_c$) together with the quarks that are known to 
occur in fundamental representations of $SU(3)_c$. 
In such a situation, anomalous values of the total angular momentum may 
be possible when the spin of a quark is combined with the angular momentum 
in the color gauge fields ({\it i.e.} gluonic degrees of freedom). 
Another possible application 
may be to the bosonic sector of the unbroken standard electroweak model. 

The possibility of obtaining half-integral angular momentum without magnetic 
charge is particularly relevant to the construction of a dual standard model 
\cite{hltv,tv,hlgstv} in which every
known quark and lepton corresponds to a (dualized) magnetic monopole.
Now, the neutrino is a spin half particle with zero electric charge and
so its magnetic counterpart would have to be a spin half object
with zero magnetic charge. What we have seen here is that such a state 
might be possible, although it is still not clear if 
the classical configuration can be (third) quantized 
as a massless spin-half particle. 

Finally we should point out the difference between the present half-integer
spin configurations and objects such as Skyrmions which can also be quantized
as fermions. In the present case, the angular momentum is calculated classically
and the half-integral value crucially depends on the 
non-Abelian nature of the gauge fields. In the case of the Skyrmion, the 
classical object does not carry half-integral
values of angular momentum. Only when the Skyrmion is quantized together with
a Wess-Zumino-Witten term, does one get the possibility of half-integral angular
momentum. A similarity, however, is that the Skyrmion needs to be
stabilized against collapse by including some higher derivative terms
in the model. In our case, the gauge configurations (such as (\ref{bpsa}))
are unstable to spreading out. If desired, they can be
stabilized by the inclusion of gravitational forces.

\

{\it Acknowledgements:} I thank Fred Goldhaber, Hong Liu, Tom Kibble,
Lawrence Krauss, John Preskill and Mark Trodden for their comments. 
This work was supported by the Department of Energy.


\begin{thebibliography}{999}

\bibitem{rjcr} R. Jackiw and C. Rebbi, Phys. Rev. Lett. {\bf 36}, 1116 (1976).

\bibitem{phth} P. Hasenfratz and G. 't Hooft, Phys. Rev. Lett. {\bf 36},
1119 (1976).

\bibitem{th} G. 't Hooft, Nucl. Phys. {\bf B79}, 276 (1974). 

\bibitem{ap} A. M. Polyakov, JETP Lett. {\bf 20}, 194 (1974).

\bibitem{ag} A. S. Goldhaber, Phys. Rev. Lett. {\bf 36}, 1122 (1976).

\bibitem{ls} J. D. Lykken and A. Strominger, Phys. Rev. Lett. {\bf 44}, 1175
(1980).

\bibitem{gh} C. Gardner and J. Harvey, Phys. Rev. Lett. {\bf 52},
879 (1984).

\bibitem{hltv} H. Liu and T. Vachaspati, Phys. Rev. {\bf D56}, 1300 (1997). 

\bibitem{taubes} C. H. Taubes, Comm. Math. Phys. {\bf 86}, 257 (1982);
{\bf 86}, 299 (1982).

\bibitem{b} E. B. Bogomol'nyi, Sov. J. Nucl. Phys. {\bf 24}, 449 (1976); 
reprinted in ``Solitons and Particles'', C. Rebbi and G. Soliani, World
Scientific, Singapore (1984).

\bibitem{ps} M. K. Prasad and C. M. Sommerfield, Phys. Rev. Lett. {\bf 35},
760 (1975).

\bibitem{mbmbtv} M. Barriola, M. Bucher and T. Vachaspati, Phys. Rev. 
{\bf D50}, 2819 (1994).

\bibitem{tv} T. Vachaspati, Phys. Rev. Lett. {\bf 76} 188 (1996).

\bibitem{hlgstv} H. Liu, G. D. Starkman and T. Vachaspati,
Phys. Rev. Lett. {\bf 78} 1223 (1997).


\end{thebibliography}
\end{document}